# Factors Influencing Change Orders in Horizontal Construction Projects: A Comparative Analysis of Unit Price and Lump Sum Contracts

Mohamed Khalafalla[1]*, Tejal Mulay[2] and Shonda L Bernadin[3]

[1]*Assistant Professor, Division of Engineering Technology, Florida A&M University, Tallahassee, Florida, USA*

[2]*Visiting Assistant Professor, Division of Engineering Technology, Florida A&M University, USA*

[3]*Associate Professor, Department of Electrical and Computer Engineering, FAMU-FSU College of Engineering, USA*

**\*Corresponding author:** Shonda L, Bernadin Assistant Professor, Division of Engineering Technology, Florida A&M University, Tallahassee, Florida, USA

## Abstract

Change orders (COs) are a common occurrence in construction projects, leading to increased costs and extended durations. Design-Bid-Build (DBB) projects, favored by state transportation agencies (STAs), often experience a higher frequency of COs compared to other project delivery methods. This study aims to identify areas of improvement to reduce CO frequency in DBB projects through a quantitative analysis. Historical bidding data from the Florida Department of Transportation (FDOT) was utilized to evaluate five factors—contracting technique, project location, type of work, project size, and duration—on specific horizontal construction projects. Two DBB contracting techniques, Unit Price (UP) and Lump Sum (LS), were evaluated using a discrete choice model. The analysis of 581 UP and 189 LS projects revealed that project size, duration, and type of work had a statistically significant influence on the frequency of change orders at a 95% confidence level. The discrete choice model showed significant improvement in identifying the appropriate contract type for a specific project compared to traditional methods used by STAs. By evaluating the contracting technique instead of project delivery methods for horizontal construction projects, the use of DBB can be enhanced, leading to reduced change orders for STAs.

## Introduction

Construction projects invariably experience changes during the design or construction phases, resulting in the need for change orders (COs) [1,2]. COs can stem from the project owner's requests for modifications to the original contract or arise from field conditions and conflicts encountered during construction [3,4]. These changes can have adverse effects on project performance, leading to increased costs and delays [5]. Researchers have suggested that the chosen project delivery method or contract type directly affects the size and frequency of these COs. Previous research suggests that the selection of project delivery methods and contract types directly influences the frequency and magnitude of COs [6-8]. Among the various project delivery methods used by State Transportation Agencies (STAs), Design-Bid-Build (DBB) is one of the most prevalent procurement techniques in the United States (9). Since more than 70 percent of STAs projects were procured through DBB, this was the method selected for the analysis of this study [10,11]. DBB involves the owner, often an STA, developing the project design internally or through the engagement of an engineer, while the construction is awarded through a separate agreement [9,12]. The DBB method can be customized like any other method to fit specific projects' needs based on the project's different administrative and management aspects [13]. The compensation for general contractors in DBB projects can take different forms, such as Unit Price (UP), Lump Sum (LS), or Cost Reimbursable contracts [11]. This study focuses on analyzing DBB projects, specifically those utilizing UP and LS contracting techniques, with the aim of understanding the factors influencing the frequency of COs. By utilizing bidding data from the Florida Department of Transportation (FDOT) between January 2015 and March 2017, the research examines the impact of contracting technique, project location, type of work, project size, and duration on the frequency of COs in horizontal construction projects. The research employs a discrete choice model to predict the frequency of COs and compare the effectiveness of UP and LS contracts in managing change orders. The findings of this research contribute to enhancing the understanding of COs.

## Background

In UP contracts, contractors are paid a fixed cost per unit of each item, such as per cubic yard of excavation or linear foot of guide rail, based on the actual measured units constructed on the project [14]. On the other hand, LS contracts involve a fixed amount of money for performing the specified work outlined in the bid documents [15,16]. LS contracts aim to simplify the payment process and reduce contract administration costs, focusing more on quality rather than individual pay items [17,18], Minchin Jr, Chini et al. [19,20]. While UP is the primary contracting technique used by State Transportation Agencies (STAs) for DBB projects, LS contracts are employed in select cases, particularly for simpler projects with well-defined scopes and minimal risk of unforeseen conditions and changes (FDOT 2017). The advantages of LS contracts include reduced contract administration costs, simplified payment processes, and a focus on quality [17,21] .How ever, managing and negotiating COs can be more challenging under LS contracts due to the absence of unit prices and difficulties in pricing and identifying changes in the scope of work [14]. Contractors may include higher contingencies in their price proposals to mitigate potential risks. Previous research has mainly focused on comparing different delivery methods such as DBB and Design-Build (DB), rather than examining different forms of DBB contract methods [22,23]. Studies have indicated that DBB projects may have a higher frequency of COs compared to DB projects [23]. Another study suggested that Lump Sum DB contracts may outperform DBB contracts in terms of schedule and cost [24]. The causes of COs in construction projects include client-related issues and a lack of national information and databases on soil conditions and services [25].

FDOT considers LS as a distinct contracting technique for simple projects and utilizes UP contracts for more complex projects (FDOT 2017). FDOT has defined ten types of COs that contractors can request during the construction phase, which have specific impacts on project constraints such as schedule and cost (FDOT 2017). The advantages of LS contracts, as identified in previous literature, include reduced contract administration costs, simplified payment processes, and greater flexibility in construction means and methods [14]. However, the disadvantages include difficulties in pricing and negotiating COs, identifying changes in the scope of work, and potentially higher contingencies in price proposals [11]. It is challenging to price COs under LS contracts due to the absence of unit prices, leading to contractors including higher contingencies in their price proposals [16]. To mitigate the disadvantages of LS contracts, project selection criteria have been developed, considering factors such as a well-defined scope of work, low risk of unforeseen conditions, low possibility of scope changes, and contractors providing the required quantities of work [19,22]. In this study, we aim to analyze DBB projects utilizing UP and LS contracting techniques, focusing on the factors influencing the frequency of COs. By examining bidding data from FDOT, we evaluate the impact of contracting technique, project location, type of work, project size, and duration on the frequency of COs in horizontal construction projects. Furthermore, we employ a discrete choice model to predict the frequency of COs and compare the effectiveness of UP and LS contracts in managing change orders.

## Methodology

The research methodology employed in this study is illustrated in Figure 1. The study began with an extensive literature review to gain insights into the contracting practices of STAs in DBB projects, as well as the processes related to COs. Relevant data and information were then collected, including historical bid data from all projects awarded by FDOT between January 2015 and March 2017. Bid tabulations published by FDOT were used to gather approximately 1,274 projects for further analysis. Among these, 705 projects were UP contracts, and 200 projects were LS contracts, which formed the focus of this study. It should be noted that the selected projects were in different life cycle stages, including executed, work begun, final acceptance, material certified, final payment authorized, and final payment made.From the collected projects, a subset was identified for in-depth analysis based on their life cycle stages. Specifically, the projects in the final payment authorization and final payment made stages were chosen, as these stages indicated that any potential changes in the construction contract had already been finalized. This subset consisted of 581 UP projects and 189 LS projects, as shown in Table 1 Additionally, provides information on the projects that included CO requests.The analysis focused on understanding the changes that occurred during the life cycle of the selected projects, with the frequency of COs requested serving as the dependent variable. The frequency of COs approved by FDOT in the analyzed projects ranged from zero to 197 COs per project, as depicted in Figure 2 For analysis purposes, the frequency of COs was categorized into three groups: low (≤10 COs), medium (11-20 COs), and high (>20 COs).

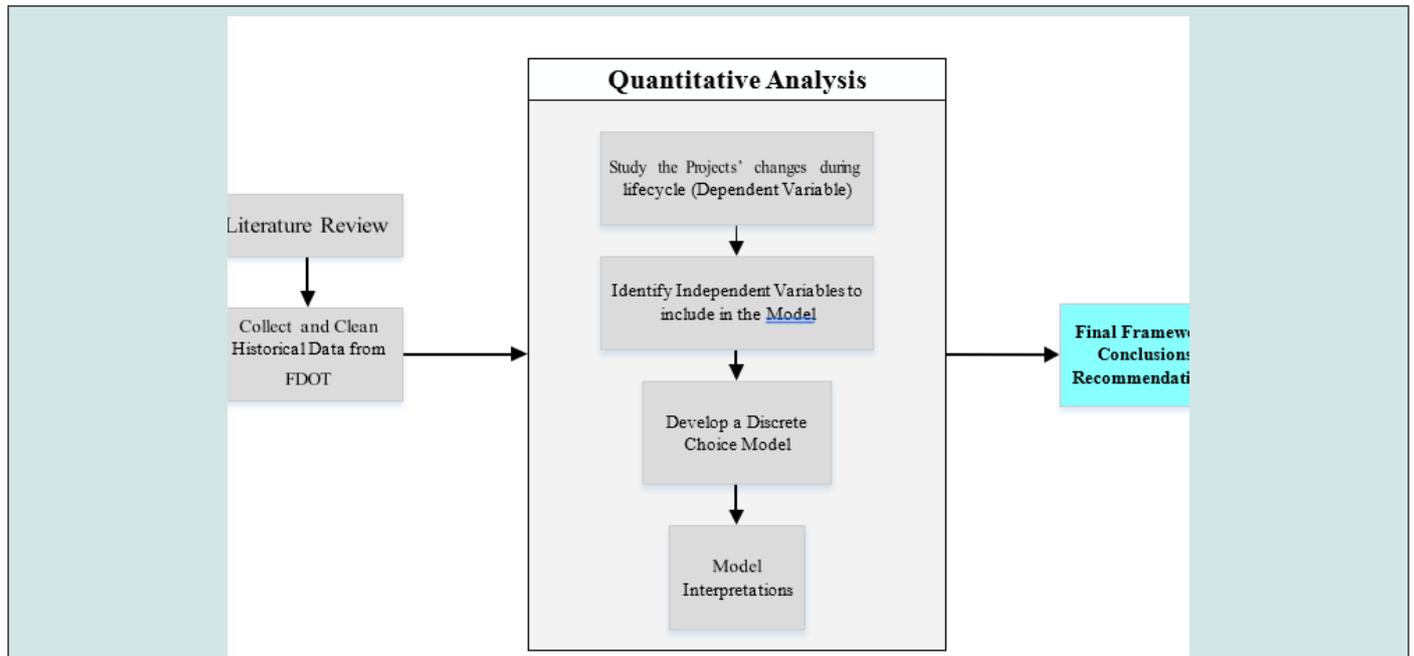

**Figure 1:** Proposed Research Methodology.

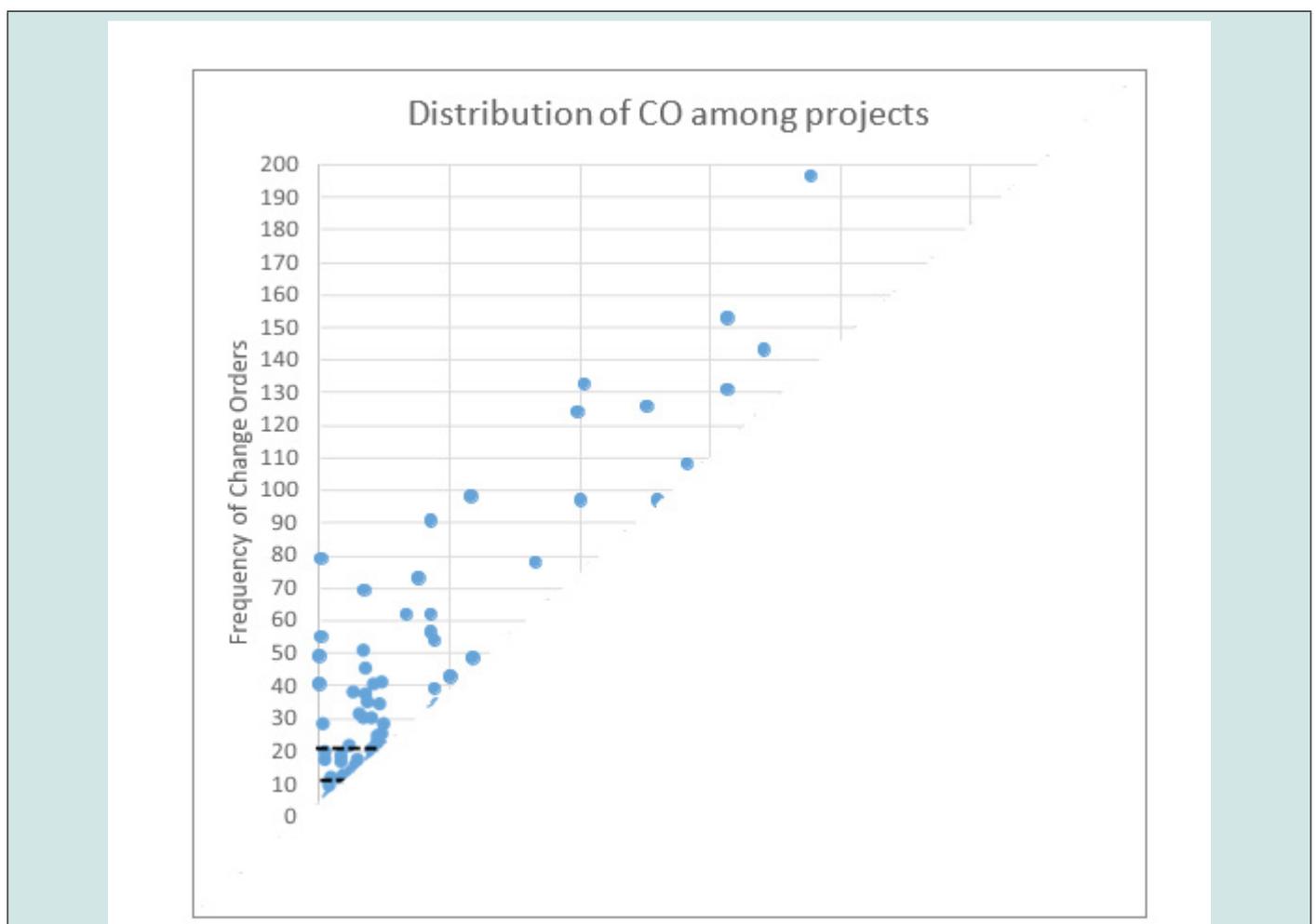

**Figure 2:** Distribution of Change Orders among Projects.

**Table 1:** The Total Number of Projects and Projects where CO Requested

| Type of Project Work | Total Number of Projects | | # Projects Requested COs | |
|---|---|---|---|---|
| | Unit Price | Lump Sum | Unit Price | Lump Sum |
| Bridge Construction | 26 | 2 | 18 | 2 |
| Bridge Repair | 95 | 4 | 35 | 1 |
| Interstate Construction (New) | 1 | 0 | 0 | - |
| Interstate Rehabilitation | 9 | 0 | 5 | - |
| Miscellaneous Construction | 50 | 30 | 29 | 9 |
| New Construction | 9 | 0 | 8 | - |
| Other | 67 | 42 | 25 | 18 |
| Reconstruction | 36 | 3 | 33 | 2 |
| Resurfacing | 229 | 82 | 125 | 35 |
| Traffic Operations | 24 | 23 | 9 | 5 |
| Widening & Resurfacing | 35 | 3 | 29 | 1 |
| Total Projects | 581 | 189 | 316 | 73 |

Several independent variables were considered in the analysis. The first variable was the contracting technique (UP or LS), as the type of contract was expected to influence the occurrence of COs. The second variable was the location of the project, which was divided into eight districts of FDOT. The third variable was the type of work, including Miscellaneous Construction, Resurfacing, Traffic Operations, and Other types. The fourth and fifth variables were the size and duration of the project, respectively, which were determined based on the agency's cost estimate and initially estimated duration before the bidding process.

**Independent Variables**

Several independent variables were considered in the analysis. The first variable was the contracting technique (UP or LS), as the type of contract was expected to influence the occurrence of COs. The second variable was the location of the project, which was divided into eight districts of FDOT. The third variable was the type of work, including Miscellaneous Construction, Resurfacing, Traffic Operations, and Other types. The fourth and fifth variables were the size and duration of the project, respectively, which were determined based on the agency's cost estimate and initially estimated duration before the bidding process.

**Model Development:**

Discrete choice models are widely used in various fields to explain and predict decision-making behavior. These models are particularly valuable in understanding consumer preferences and choices in market research, where they help analyze consumer demand and address pricing and demand estimation challenges [26,27]. In transportation, discrete choice models play a crucial role in predicting the demand for transportation systems [28]. These models can be categorized into binomial and multinomial choice models, depending on the number of alternatives within the dependent variable (Table 2). Binomial models are employed when there are only two alternatives to choose from, while multinomial choice models are utilized when three or more alternatives exist. To account for errors in the model, a logit or probit distribution function can be incorporated. In this study, a logit ordered multinomial model was adopted. This choice was motivated by the nature of the dependent variable (frequency of COs), which involves multiple outcome variables that require a rank order. Considering these characteristics, using a logit error distribution function and employing ordinal regression techniques were deemed appropriate for the analysis. A discrete choice model, specifically a logit ordered multinomial model, was developed to analyze and predict the behavior of selecting between the different alternatives (dependent variable) based on observed characteristics (independent variables). The model was calibrated and evaluated using IBM SPSS Statistics 25® software, and the parameter estimates were obtained. These estimates were used to interpret the significance and impact of each independent variable on the frequency of COs.

**Table 2:** Case Processing Summary

| | | N | Percentage of Total Data |
|---|---|---|---|
| Category of the frequency of change orders | Low | 388 | 57.60% |
| | Medium | 123 | 18.20% |
| | High | 163 | 24.20% |
| Contracting Technique | Lump Sum | 176 | 26.10% |
| | Unit Price | 498 | 73.90% |
| Type of work | Resurfacing | 274 | 40.70% |
| | Traffic Operations | 39 | 5.80% |
| | Miscellaneous Construction | 70 | 10.40% |
| | Other Types | 291 | 43.20% |
| Florida District | District 1 - Southwest Florida | 99 | 14.70% |
| | District 2 - Northeast Florida | 104 | 15.40% |
| | District 3 - Northwest Florida | 122 | 18.10% |
| | District 4 - Southeast Florida | 62 | 9.20% |
| | District 5 - Central Florida | 85 | 12.60% |
| | District 6 - South Florida | 96 | 14.20% |
| | District 7 - West Central Florida | 90 | 13.40% |
| | District 8 - Florida's Turnpike | 16 | 2.40% |
| Valid | | 674 | 100.00% |
| Missing | | 96 | |
| Total | | 770 | |

## Results

The multinomial ordinal logit regression model was employed using IBM SPSS Statistics 25® to analyze the data with a confidence level of 95 percent (p = 0.05). Table 3 provides an overview of the case processing involved in the analysis. Models with log-likelihood values close to zero are considered to be the best-fit models. In Table 4, the final new model has a log-likelihood value of 958.63, which is closer to zero compared to the intercept-only baseline model's value of 1309.73. This indicates that the new model better captures the variance in the outcome and represents an improvement over the original model. The chi-square value of 351.1 indicates that the data is significant, supporting the conclusion that the new model is an improvement. The analysis identified several variables that were not statistically significant at a 95 percent confidence level (p = 0.05). With the exception of the size of the project and the type of project (resurfacing), most of the variables were found to be statistically insignificant (Table 5).

**Table 3:** Model Fitting Information

| Model | -2 Log-Likelihood | Chi- Square | Degree of Freedom | P-Value |
|---|---|---|---|---|
| Intercept Only | 1309.734 | | | |
| Final | 958.63 | 351.104 | 13 | 0 |

**Table 4:** Parameter Estimates

| | Variables | Coefficient Estimate | P-Value |
|---|---|---|---|
| Threshold | Low | 2.454 | 0.001 |
| | Medium | 3.813 | 0 |
| Location | Size of the Project | 1.34E-07 | 0 |
| | Duration of the Project | 0.009 | 0 |
| | **Contracting Technique** | | |
| | Ø Base: Unit Price | | |
| | Ø Lump Sum | -0.503 | 0.037 |
| | **Type of work** | | |
| | Ø Base: Other Types | | |
| | Ø Resurfacing | 0.895 | 0 |
| | Ø Traffic Operations | 0.401 | 0.378 |
| | Ø Miscellaneous Construction | 0.892 | 0.007 |
| | **Florida District:** | | |
| | Ø Base: District 8 - Florida's Turnpike | | |
| | Ø District 1 - Southwest Florida | -0.295 | 0.678 |
| | Ø District 2 - Northeast Florida | -1.188 | 0.093 |
| | Ø District 3 - Northwest Florida | -0.349 | 0.62 |
| | Ø District 4 - Southeast Florida | -0.092 | 0.899 |
| | Ø District 5 - Central Florida | -0.464 | 0.523 |
| | Ø District 6 - South Florida | -0.999 | 0.163 |
| | Ø District 7 - West Central Florida | -0.413 | 0.566 |

Table 5: Interpretations from the Model

| Variable | Significant | Meaning |
|---|---|---|
| Size of the Project | Yes | As the size of the project increases, the number of change orders is more likely to increase. |
| Duration of the Project | Yes | As the duration of the project increases, the number of change orders is more likely to increase. |
| **Contracting Technique** | | |
| **Ø Base**: Unit Price | | |
| Ø Lump-Sum | Yes | Projects procured through LS are less likely to have COs than Projects procured through UP. |
| **Type of work** | | |
| **Base**: Other Types | | |
| Ø Resurfacing | Yes | Resurfacing projects are more likely to have more COs than other types of projects |
| Ø Traffic Operations | No | Traffic Operation projects are more likely to have more COs than other types of projects |
| Ø Miscellaneous Construction | Yes | Miscellaneous Construction projects are more likely to have more COs than other types of projects |
| **Florida Location** | | |
| **Base**: District 8 | | |
| Ø District 1 | No | Projects in District 1 are less likely to have more COs than projects in District 8 |
| Ø District 2 | No | Projects in District 2 are less likely to have more COs than projects in District 8 |
| Ø District 3 | No | Projects in District 2 are less likely to have more COs than projects in District 8 |
| Ø District 4 | No | Projects in District 4 are less likely to have more COs than projects in District 8 |
| Ø District 5 | No | Projects in District 5 are less likely to have more COs than projects in District 8 |
| Ø District 6 | No | Projects in District 6 are less likely to have more COs than projects in District 8 |
| Ø District 7 | No | Projects in District 7 are less likely to have more COs than projects in District 8 |

## Conclusions And Recommendations

Construction projects often undergo changes in scope, time, or cost during their life cycle, which are managed through the issuance of change orders (COs) by owners. The frequency of COs requested can significantly impact the expected outcomes of projects, with lower frequencies generally leading to better project outcomes. This study aimed to identify the factors that contribute to COs and their effect on the frequency of COs approved for horizontal construction projects. The analysis utilized a multinomial ordinal regression discrete choice model and focused on completed projects undertaken by FDOT between 2015 and 2017, consisting of 581 unit price (UP) and 189 lump sum (LS) contracts. The results of the analysis indicate that project size, project duration, and contracting technique (UP or LS) were statistically significant in influencing the frequency of change orders during the construction life of a project. Specifically, as the size and duration of a project increases, the likelihood of experiencing more COs also increases. Projects procured through the LS contracting technique were found to have a lower likelihood of COs compared to projects procured through UP. Moreover, the data reported for resurfacing and miscellaneous construction projects demonstrated a higher likelihood of having more COs compared to other types of project work. For future research, it is suggested to explore additional factors, such as studying the influence of CO size as a percentage of the project's cost. Developing an objective approach for selecting projects suited for the LS contracting technique, rather than relying solely on expert judgment, could also be beneficial. Furthermore, future research could focus on developing a framework for selecting projects to be procured through the LS technique. By expanding the understanding of these factors, construction stakeholders can make more informed decisions to minimize the occurrence of COs and enhance project performance.

## References

1. Moselhi O, Leonard C, Fazio P (1991) Impact of change orders on construction productivity. Canadian journal of civil engineering 18(3): 484-492.

2. Ahmed S, Arocho I (2021) Analysis of cost comparison and effects of change orders during construction: Study of a mass timber and a concrete building project. Journal of building engineering 33: 101856.

3. Riley DR, Diller BE, Kerr D (2005) Effects of delivery systems on change order size and frequency in mechanical construction." Journal of Construction Engineering and Management 131(9): 953-962.

4. Al Malki Y M, Alam MS (2021) Construction claims their types and causes in the private construction industry in the Kingdom of Bahrain." Asian Journal of Civil Engineering 22(3): 477-484.

5. Hanna A S, Camlic R, Peterson PA, Nordheim EV (2002) Quantitative definition of projects impacted by change orders. Journal of Construction Engineering and Management 128(1): 57-64.

6. Pakalapati K, Khalafalla M, Rueda-Benavides J (2020) Using moving-window cross-validation algorithm to optimize bid-based cost estimating data usage. International Journal of Construction Management p. 1-9.

7. Rueda Benavides J, Khalafalla M, Miller M, Gransberg D (2022) Cross-asset prioritization model for transportation projects using multi-attribute utility theory: a case study." International Journal of Construction Management p. 1-10.

8. Rueda-Benavides J, Gransberg D, Khalafalla M, Mayorga C (2023) Probabilistic cost- based decision-making matrix: IDIQ vs. DBB contracting. Construction Management and Economics p. 1-15.

9. Khalafalla M, Rueda JA (2020) Methodology to assess the impact of lump-sum compensation provisions on project schedules. Journal of Management in Engineering 36(4).

10. Ibbs C W, Kwak YH, Ng T, Odabasi AM (2003) Project delivery systems and project change: Quantitative analysis. Journal of Construction Engineering and Management 129(4): 382-387.

11. Khalafalla MK M (2019) Cost-duration-based lump sum project selection framework using stochastic methods for design-bid-build resurfacing projects (Doctoral dissertation, Auburn University.

12. Gransberg DD, Shane JS (2010) Construction manager-at-risk project delivery for highway programs, Transportation Research Board.

13. Walewski J, Gibson GE, Jasper J (2001) Project delivery methods and contracting approaches available for implementation by the Texas Department of Transportation, Center for Transportation Research, Bureau of Engineering Research, University of Texas at Austin.

14. Khalafalla M, Rueda Benavides J (2018) Unit Price or Lump Sum? A Stochastic Cost- Based Decision-Making Tool for Design-Bid-Build Projects. Transportation Research Record.

15. Clough R, Sears G (1981) Construction Contracting John Willy and sons Inc. New York.

16. Minchin Jr, Chini RE, Ptschelinzew AR, Shah L, Zhang D, et al. (2016) Alternative contracting research.

17. Caltrans CD, oT (2007) Innovative Procurement Practices. Innovative Procurement and Contracting Methods. C. D. o. Transportation.

18. Ellis Jr R D, Pyeon JH, Herbsman ZJ, Minchin E, Molenaar K, et al. (2007) Evaluation of alternative contracting techniques on FDOT construction projects.

19. FDOT, F D oT (2017) Plans Preparation Manual. Lump Sum Project Guidelines. F. D. o.Transportation, FDOT. 1.

20. Park J, Kwak YH (2017) Design-Bid-Build (DBB) vs. Design-Build (DB) in the US public transportation projects: The choice and consequences. International Journal of Project Management 35(3): 280-295.

21. DOT, PF (2003) Lump Sum Project Guidelines. T. P. F. S. o. Alaska), Alaska Department of Transportation and Public Facilities.

22. DOT, PF, A D, o T a P F (2003) Lump Sum Project Guidelines Alaska Department of Transportation and Public Facilities.

23. Shrestha PP, O'Connor JT, Gibson Jr GE (2011) Performance comparison of large design-build and design-bid-build highway projects. Journal of Construction Engineering and Management 138(1): 1-13.

24. Fathi M, Shrestha PP, Shakya B (2020) Change orders and schedule performance of design-build infrastructure projects: Comparison between highway and water and wastewater projects. Journal of Legal Affairs and Dispute Resolution in Engineering and Construction 12(1): 04519043.

25. Train K E (2009) Discrete choice methods with simulation, Cambridge university press.

26. Greene WH (2008) Discrete Choice Modeling. Teoksessa T Mills, K Patterson (toim.) The Palgrave Handbook of Econometrics Applied Econometrics p. 2.

27. Alnuaimi A S, Taha RA, Al Mohsin M, Al-Harthi AS (2009) Causes effects benefits and remedies of change orders on public construction projects in Oman. Journal of Construction Engineering and Management 136(5): 615-622.

28. Antonini G, Bierlaire M, Weber M (2006) Discrete choice models of pedestrian walking behavior. Transportation Research Part B: Methodological 40(8): 667-687.

**Citation:** Mohamed Khalafalla*, Tejal Mulay and Shonda L Bernadin. Factors Influencing Change Orders in Horizontal Construction Projects: A Comparative Analysis of Unit Price and Lump Sum Contracts. Tr Civil Eng & Arch 4(4)- 2022. TCEIA.MS.ID.000192.
DOI: 10.32474/TCEIA.2023.04.000192